\documentclass[twoside,a4paper,11pt]{sea10}
\usepackage{graphicx}
\usepackage{hyperref}
\begin{document}
\pagenumbering{arabic}
\pagestyle{myheadings}
\thispagestyle{empty}
{\flushleft\includegraphics[width=\textwidth,bb=58 650 590 680]{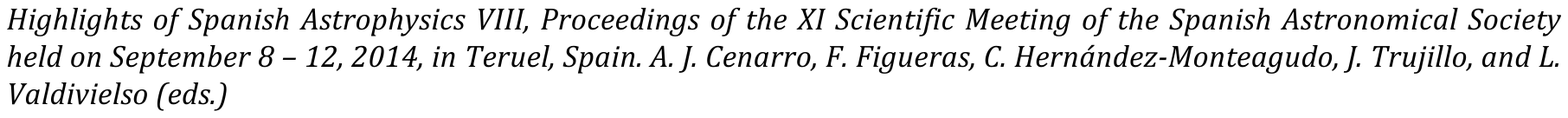}}
\vspace*{0.2cm}
\begin{flushleft}
{\bf {\LARGE
%
%%% TITLE of the paper. 
%%% TITLE of the paper. 
Discordant optical and X-ray classification of AGN
%
% Do not delete next few lines
}\\
\vspace*{1cm}
%
%%% Include here the LIST OF AUTHORS.
%%% Include here the LIST OF AUTHORS.
%%% Note that the last author has to be preceeded by an AND.
Ordov\'as-Pascual, I.$^{1}$,
Mateos, S.$^{1}$,
Carrera, F. J.$^{1}$, 
and 
Wiersema, K.$^{2}$
%
% Do not delete next few lines
}\\
\vspace*{0.5cm}
%
%%% AFFILIATIONS LIST.
%%% and the AFFILIATIONS LIST. Note that one affiliation per line.
%%% Add as many affiliations as necessary. 
$^{1}$
Instituto de Física de Cantabria (CSIC-UC) 39005 Santander, Spain\\
$^{2}$
Department of Physics and Astronomy, University of Leicester, Leicester, UK, LE1 7RH
%
% Do not delete next few lines
\end{flushleft}
%
% Headings
\markboth{
%%% Type the SHORT version of the paper title.
%%% Type the SHORT version of the paper title.
Discordant optical and X-ray classification of AGN
}{ % Do not delete
%
%%%  First Author \& Second Author   OR   First-author et al. 
%%%  First Author \& Second Author   OR   First-author et al. if the author list 
%%% contains three or more authors.
 Ordov\'as-Pascual et al. 
% 
% Do not delete next few lines
}
\thispagestyle{empty}
\vspace*{0.4cm}
\begin{minipage}[l]{0.09\textwidth}
\ 
\end{minipage}
\begin{minipage}[r]{0.9\textwidth}
\vspace{1cm}
\section*{Abstract}{\small
%
% ABSTRACT ABSTRACT ABSTRACT
% ABSTRACT ABSTRACT ABSTRACT
%%% Type the ABSTRACT of your paper

To provide insight into the apparent mismatch between the optical and X-ray absorption properties observed in 10-30 \% of Active Galactic Nuclei (AGN), we have conducted a detailed study of two X-ray unabsorbed AGN with a type-2 optical spectroscopic classification. In addition to high quality X-ray spectroscopic observations, that we used to determine both the AGN luminosities and absorption, we have a VLT/XSHOOTER UV-to-near-IR high resolution spectrum for each object, that we used to determine the AGN intrinsic emision corrected for both contamination from the AGN hosts and extinction. Our analysis has revealed that the apparent mismatch is provoked by galaxy dilution. We dilution of two AGN with extreme properties: one of them has an intrinsically very high Balmer decrement while the other lies in a galaxy more massive than expected. 
%
% Do not delete next few lines
\normalsize}
\end{minipage}

\section{Introduction \label{intro}}

According to the unified model of AGN, type-1 and type-2 AGN are the same kind of object. The differences in their observed properties can be explained by orientation effects. If our line-of-sight to the central engine intercepts the dusty torus invoked by unified models, the UV/optical broad emission lines are obscured, and the X-ray emission is absorbed. In this case, the AGN is classified as type-2. On the other hand, in type-1 we have a direct view of the central engine, the broad lines from the Broad Line Region (BLR) are present in the optical spectrum, and the X-ray emission is unobscured.

The classification of AGN using either the optical range or X-rays should agree according to this model, but approximately, 10-23\% of type-1 AGN present an X-ray absorbed spectrum, and 3-17\% of type-2 AGN are X-ray unabsorbed (\cite{mateos05}, \cite{corral11}, \cite{merloni14}). 

The objective of this study is to uncover the apparent mismatch between the UV/optical and X-ray properties of X-ray unobsured type-2 AGN. We tested three possible scenarios: a) the objects are Compton-Thick; ie. the nuclear N$_H$ column density is equal to or larger than the inverse of the Thomson cross-section (N$_H>\sigma_{T}^{-1}$=1.5$\times$10$^{24}$ cm$^{-2}$). In this case, the X-ray emission below 10 keV is totally obscured and the spectrum would be scattered and reflected emission, a strong Fe line at 6.4-7 keV; b) the AGN signatures, the broad UV/optical emission lines, are present but diluted by the host galaxy contribution; c) the mismatch is provoked by intrinsic non-standard properties of the AGN, such as a weak BLR emission, and/or their hosts.

\section{The sample \label{sample}}

The two AGN analysed in this work were selected from the Bright Ultra-hard XMM-Newton Survey (BUXS). This is an X-ray selected sample at 4.5-10 keV energies that includes 258 AGN with (4.5-10 keV) $>$ 6$\times 10^{-14}$ erg s$^{-1}$ cm$^{-2}$ detected in a total sky area of 44.43 deg$^{2}$. The optical spectroscopic completeness is $>$98\% \cite{mateos12}. We have selected 2 from this sample with an apparent mismatch between the UV/optical and X-ray properties. 

\begin{figure}
\center
\includegraphics[width=1\textwidth]{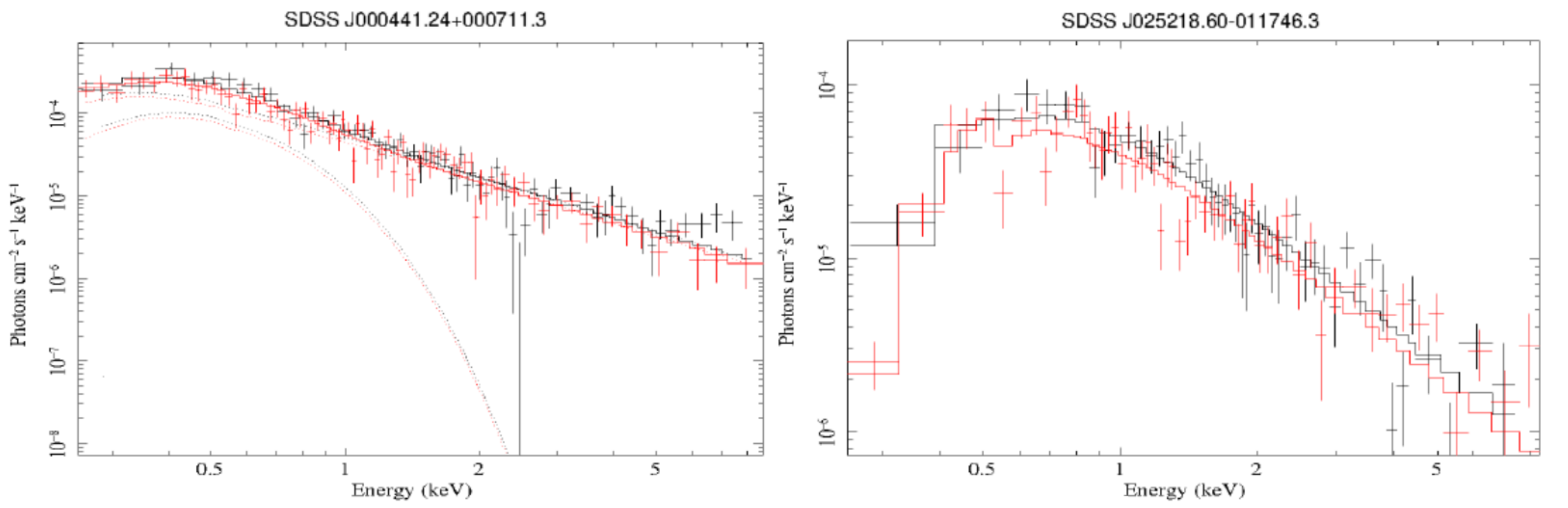} 
\caption{\label{muestrai} XMM-Newton spectra for the selected objects. The 2-10 KeV luminosities are, in log(L$_{2-10 KeV}$ erg s$^{-1}$) 42.76 for J000441.24+000711.3 and 41.25 for J025218.60-011746.3. For J000441.24+000711.3 using a best-fit model bb+po (black-body + power law) we obtained an upper limit for the X-ray column density of $<$6.7$\times$ 10$^{20}$ cm$^{-2}$, while for J025218.60-011746.3 apo (absorbed powe law) we obtained 1.7$_{-1.4}^{+2.0}\times$10$^{21}$ cm$^{-2}$. The first object is at z=0.1077 and the latter at z=0.0246.
}
\end{figure}

SDSS J000441.24+000711.3: it is X-ray unobscured. The SDSS public spectrum show a broad component in H$_\alpha$ but not in H$_\beta$, so it is a Seyfert 1.9.
SDSS J025218.60-011746.3: is a Sb edge-on galaxy with low X-ray obscuration. 6dF public spectrum indicates a Seyfert 2 classification. 
In Figrure \ref{muestrai} it is shown the XMM-Newton spectra for the selected objects and in the caption we list their main properties.

\section{The Data \label{xsh}}

We present high resolution spectra for both objects from VLT/XSHOOTER \cite{vernet11}. The observations were taken with a 1.0''x11'' slit for the arm (UVB) and a 0.9''x11'' slit for visible and near-infrared arms (VIS and NIR, respectively). The wavelength coverage is 3000-5500 \AA ~for the UVB, 5500-10000 \AA ~for VIS and 10000-25000 \AA ~for NIR. For each frame and for both objects, the exposition time was 1420 seconds for UVB and VIs and 480 seconds for NIR. Full details on the reduction of the data will be presented in Ordov\'as-Pascual et al. 2015 (in preparation). In both cases we used a 2'' region to extract the AGN spectra. We also obtained the spectra from the AGN host galaxies by extracting spectra 1'' away from the centre for J000441.24+000711.3 and 1.25'' for J025218.60-011746.3 using a 1.5'' aperture.

\section{Spectral analysis \label{analysis}}

The XSHOOTER spectra are dominated by the host galaxy enmission, and most AGN signatures are completely diluted.

%After extracting the spectra from the central regions of the sources and the host galaxies, it is clear that the continuum is similar, so the stellar contamination could cause the discrepancy of the classifications. If the star light dominates the spectrum, the AGN singatures could be hidden in the spectrum. 

\begin{figure}
\center
\includegraphics[width=1\textwidth]{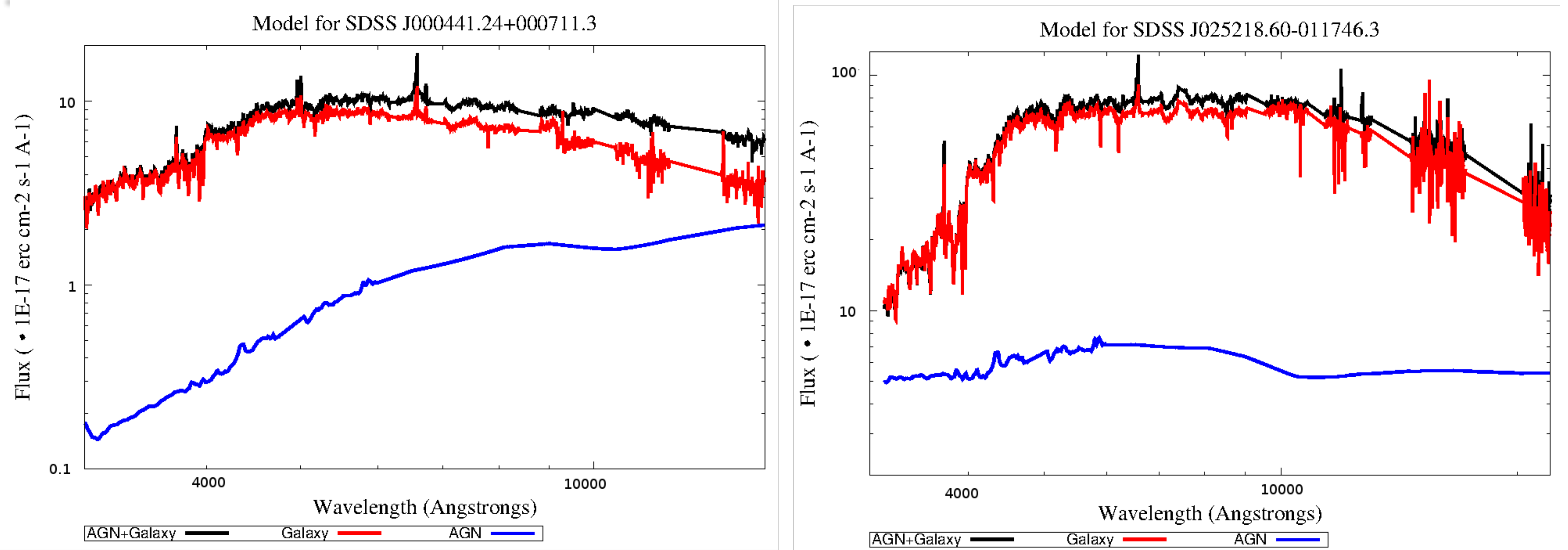} 
\caption{\label{fig1} Decomposition of the XSHOOTER spectra in host galaxy contribution and obscured AGN.
}
\end{figure}

The next step is to obtain the relative contribution of the AGN and the galaxy. This is conducted using the CIAO's SHERPA fitting tool \cite{freeman01}. We used a compound model of the extracted galaxy spectrum, a QSO1 template \cite{polletta07} with the nuclear extinction as a free parameter. The QSO1 model is corrected for slit loses. We fit the nuclei emission using a gaussian profile in the adquisition image, and then we compute the fraction of this profile that enters through the slit. We used an exctinction model of the Small Magellanic Cloud (SMC; \cite{gordon03}, \cite{allen}). The spectra of the AGN host galaxies were recalibrated to match in flux with the  Ca II K ($\lambda$-rest-frame 3933 \AA) and Ca II H ($\lambda$-rest-frame 3968 \AA) galactic absorption lines. In Table \ref{tab} we list the UV/optical derived extinction.  Fig. \ref{fig1} shows the result. After correcting for both host galaxy contamination and extinction, both objects show a broad lines in H$_{\alpha}$, and also J000441.24+000711.3 show a broad line in H$_\beta$-. 

\section{Properties of the SMBHs and their hosts \label{mass}} 

\begin{figure}
\center
\includegraphics[width=1\textwidth]{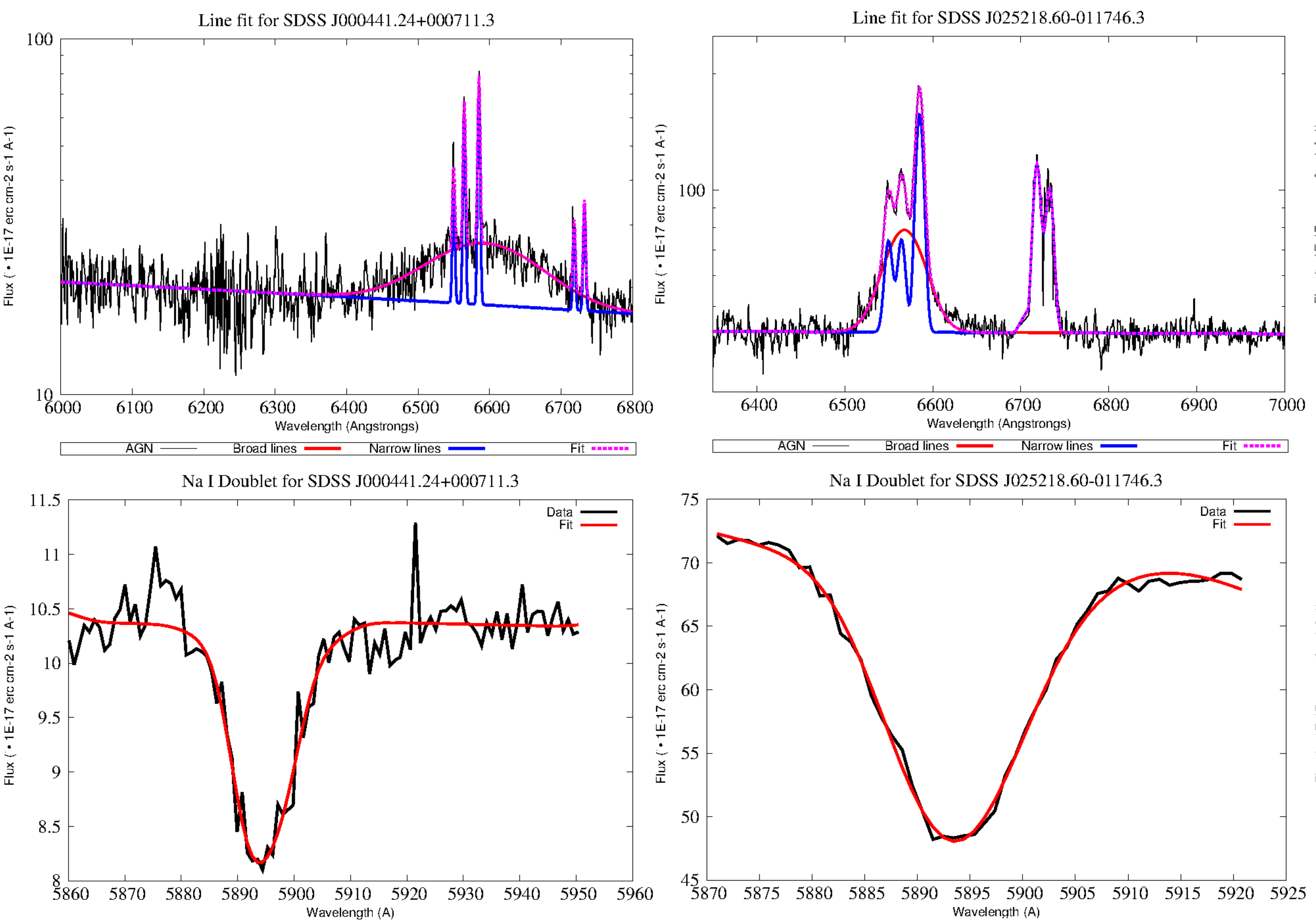} 
\caption{\label{fig2} Top: Spectra of the intrinsic AGN emission with the decomposition of the components (continuum, broad an narrow emision lines) in the zone of H$_{\alpha}$. Bottom: Spectra and fit of the NaI doublet galactic absorption lines.
}
\end{figure}

We have analyzed the zone of H$_{\alpha}$ to determine the mass of the SMBHs. The continuum is fitted by a straight line. The broad an narrow emission lines are fitted using gaussian profiles. The line widths have been corrected for the instrumental resolution of XSHOOTER. In Fig. \ref{fig2} we show the results from this analysis. To compute the M$_{SMBH}$ we used the expresion in \cite{xiao11}. The measurements for the M$_{SMBH}$ are listed in Table \ref{tab}.

The M-$\sigma$ relation consist in a linear correlation of the mass of a SMBH with the stellar velocity dispersion of the elliptical galaxy or bulge (\cite{park12}, \cite{xiao11}). The doublet of NaI is an absorption feature that it can be used to calculate the line-of-sight velocity dispersion (LOSVD) through a deconvolution of the lines. In this case we use the templates from \cite{bruzual03}. We adjust the NaId with a convolution of the chosen template and a gaussian function. The standard dispersion of this gaussian function gives the desired LOSVD. In Fig. \ref{fig2} we show absorption lines and the results of the fit. The value obtained by the fitwere also corrected for the instrumental resolution of XSHOOTER. We calculate the mass of the galaxy using  \cite{cappellari06}.

\begin{table}[ht]
\center
\caption{\label{tab} Extiction, SMBH mass and dynamical mass of the AGN host galaxies.} 
\begin{tabular}[width=1\textwidth]{ccccc} 
\hline 
Object  & A(V) (mag) & $log\left(\frac{M_{BH}}{M_{\odot}} \right)$ & $log\left(\frac{M_{dyn}}{M_{\odot}} \right)$ & $\sigma$  (km s$^{-1}$) \\ [0.5ex]   % [0.5ex] adds vertical space 
\hline
J000441.24+000711.3  & 0.03080$\pm$0.00075 & 8.21$^{+0.12}_{-0.39}$ & 11.07$^{+0.21}_{-0.41}$ & 157$\pm$48 \\ % inserting body of the table
J025218.60-011746.3  & 0.01518$\pm$0.00014 & 6.51$^{+0.11}_{-0.44}$ & 10.85$^{+0.05}_{-0.06}$ & 271$\pm$15 \\ [1ex]  % [1ex] adds vertical space 
\hline
\end{tabular} 
%\caption{ Results from the spectral analysis of the XSHOOTER data and emission line fits. 
%}

\end{table}

\section{Discussion \label{discuss}}

In this work we have investigated the origin of the apparent mismatch of the optical and X-ray classifications of two X-ray unpbscured type-2 AGN by obtaining the intrinsic emission of the AGN.We have investigated three possible explanations proposed in Section \ref{intro}.

The values of the N$_{H}$ column derived from the X-ray spectroscopy indicate that none of the AGN are Compton-thick. The X-ray spectrum is too steep to be dominated by scattered or reflected emission, and there is no sign of a strong reflection Fe line. This is supported by the L$_{2-10~keV}$/[OIII] ratio  (being [OIII] the unreddened luminosity of the  [OIII] emission line at $\lambda$-rest-frame 5007 \AA) \cite{bassani99}, that in the case of Compton-thick sources is lower than 1. This ratio is 37 for J025218.60-011746.3 and 385 for J000441.24+000711.3.

In both objects, there is significant host dilution of the nuclear emission in the UV-optical range. In the UV range the AGN contribution is in the same order as the uncertainly of the XSHOOTER spectra, and its contribution begins to be significant in the optical and near infrarred wavelengths. Only by subtracting the nuclear emission from the host galaxy contribution it is possible to uncover the broad H$_{\alpha}$ contribution, which was partially hidden in J000441.24+000711.3 and totally hidden in J025218.60-011746.3.

J000441.24+000711.3 on one hand follows the M-$\sigma$ relation. The dynamical mass of the galaxy is the expected in comparison with the M$_{SMBH}$ as well. Using the AGN emission after subtracting the stellar emission but without correcting for extinction we can calculate the Balmer decrement of the broad emission lines, and then we can calculate the N$_H$ column derived from this Balmer decrement as in \cite{carrera04}. The Balmer decrement of the obscured broad lines (H$_\alpha$/H$_\beta$=6.86$_{-3.77}^{+2.90}$) gives a higher N$_H$ column than the one obtained in Section \ref{analysis} and the X-ray derived one (5.0$_{-3.44}^{+1.52} \cdot $10$^{21}$ cm$^{-2}$ versus 4.0$ \cdot $10$^{20}$ cm$^{-2}$ and $<$ 6.0$ \cdot $10$^{20}$ cm$^{-2}$ respectively), assuming an intrinsic Balmer decrement of 2.11 \cite{jin12}. An intrinsic Balmer decrement of 6.25$_{-3.43}^{+2.65}$ would be needed in order to obtain the same N$_H$ column as in \ref{analysis}. Geometrically, it is impossible to find a way to obscure the BLR one factor higher than the accretion disk (where the AGN continuum is originated) and the corona (where the X-ray emission is originated). We conclude that this low Balmer decrement is an intrinsic property of this AGN. This is why the broad components are diluted in the galactic star light and the mismatch is originated in this object. 

J025218.60-011746.3 falls below the M-$\sigma$ relation (falls in the 2$\sigma$-3$\sigma$ range), the mass of the bulge is higher than expected compared to the SMBH mass [9], meaning that the star light contribution outshines the broad components of the nuclear emission. We do not found a broad component of H$_\beta$.

In conclusion, the AGN signatures are hidden because the galaxy contribution outshines them for two different reasons. The complete study will be presented in Ordov\'as-Pascual et al. 2015 (in preparation).

\section*{Acknowledgments}   % Do not delete if you declare acknowledgments
IO-P, SM and FJC acknowledge funding by the spanish Plan Nacional grant AYA2012-31447. SM and FJC acknowledge financial support from the ARCHES project (7th Framework of the European Union, No. 313146)

%This research has made use of the NASA/IPAC Extragalactic Database (NED) which is operated by JPL, Caltech, under contract with the National Aeronautics and Space Administration. The CASSIS is a product of the Infrared Science Center at Cornell University, supported by NASA and JPL. 
%
% Do not delete the next few lines

%
\end{document}